Article
# Black Holes Immersed in Galactic Dark Matter Halo

Alexey Dubinsky

University of Seville, 41004 Seville, Spain; dubinsky@ukr.net

**Abstract:** We analyze the quasinormal modes (QNMs) of scalar, electromagnetic, and Dirac test fields in the background of a black hole immersed in a galactic dark matter halo. The analytic black hole solution considered here is sourced by a physically motivated halo density profile that leads to a flat galactic rotation curve. Using the sixth-order WKB method with Padé approximants, we compute the QNM spectra for various field spins and parameter values, and provide numerical data in tabulated form. In addition to the numerical analysis, we derive analytic expressions for the quasinormal frequencies in the eikonal limit and beyond, by means of an expansion in inverse powers of the multipole number. We also calculate the Unruh temperature perceived by a static observer in the halo-modified spacetime. Our results demonstrate that the presence of the dark matter halo leads to observable modifications in the QNM spectra only if the density or compactness of the galactic halo is extraordinary high, so that quasinormal ringing a reliable observable for testing the black hole geometry, even in the presence of galactic environments.

## 1. Introduction

Black holes are not isolated objects in nature: they are typically embedded in astrophysical environments, including accretion disks, magnetic fields, and dark matter halos. Understanding how these environments affect observable quantities is essential, especially in the era of precision gravitational wave astronomy. One of the key signatures of black holes is the spectrum of their quasinormal modes (QNMs) [1–3], which govern the late-time behavior of perturbations and characterize the ringdown phase of a merger.

Quasinormal modes are complex frequencies that satisfy purely ingoing boundary conditions at the black hole horizon and purely outgoing conditions at spatial infinity (or a cosmological horizon). They depend only on the background geometry and the spin of the perturbing field, and not on the initial data, making them excellent probes of the spacetime itself. In recent years, there has been growing interest in understanding how QNMs are modified in alternative theories of gravity and in the presence of matter distributions.

In this work, we investigate QNMs in a black hole spacetime influenced by a surrounding galactic halo. The halo profile is chosen to yield a flat rotation curve at large distances, consistent with observational constraints on dark matter. The resulting metric is asymptotically flat and reduces to the Schwarzschild solution when the halo parameters vanish.

We consider test field perturbations of three types: scalar (spin 0), electromagnetic (spin 1), and Dirac (spin 1/2). For each case, we compute the QNMs using the WKB method with Padé resummation, which is known to provide high accuracy for low-lying modes. We supplement the numerical data with analytic approximations obtained in the eikonal regime and its subleading corrections, based on inverse multipole expansions. In addition, we compute the local Unruh temperature perceived by a static observer, generalizing the notion of Hawking radiation to finite-radius observers in this spacetime. It is worth of mentioning that various radiation phenomena, including quasinormal modes, have broadly discussed for supermassive black holes in other configurations of galactic halo [4–12].

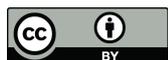

The structure of the paper is as follows. In Section 2, we review the analytic black hole solution describing a spherically symmetric galactic halo with a flat rotation curve and derive the metric. In Section 3, we formulate the perturbation equations for test scalar, electromagnetic, and Dirac fields, define the effective potentials, and describe the WKB method with Padé resummation used to compute quasinormal frequencies. In Section 4, we present analytic expressions for quasinormal modes in the eikonal and beyond-eikonal regimes, including explicit formulas for the field-specific corrections. Section 5 is devoted to the Unruh temperature as measured by a static observer in the halo-modified spacetime. We summarize our findings and discuss their implications in Section 6.

## 2. Black Hole in a Dark Matter Halo: An Analytic Solution

In this section, we consider a black hole solution embedded in a spherically symmetric galactic dark matter halo, derived in [13, 14] as an analytic extension of the Schwarzschild metric. The halo is modeled by a physically motivated dark matter density profile that reproduces the flat rotation curves of galaxies [15]:

$$\rho(r) = \frac{V_c^2}{4\pi G} \cdot \frac{3a^2 + r^2}{(a^2 + r^2)^2}, \tag{1}$$

where $V_c$ is the asymptotic circular velocity and $a$ is the core radius of the halo.

To derive a closed-form analytic expression for the spacetime geometry, the authors start from a metric ansatz of the form:

$$ds^2 = -\left[1 + F(r)\right] dt^2 + \left[1 + F(r)\right]^{-1} dr^2 + r^2 d\Omega^2, \tag{2}$$

where the function $F(r)$ is to be determined by solving the Einstein field equations with the density profile (1) as the matter source.

By integrating the field equations, one finds that:

$$F(r) = \frac{C_1}{r} + \frac{2V_c^2 r^2}{a^2 + r^2}, \tag{3}$$

where $C_1$ is an integration constant. To ensure compatibility with the Schwarzschild solution in the limit $V_c \to 0$, the authors fix the integration constant as $C_1 = -2M$.

Substituting this result into the metric ansatz (2), the final form of the solution is:

$$ds^2 = -\left[1 - \frac{2M}{r} + \frac{2V_c^2 r^2}{a^2 + r^2}\right] dt^2 + \left[1 - \frac{2M}{r} + \frac{2V_c^2 r^2}{a^2 + r^2}\right]^{-1} dr^2 + r^2 d\Omega^2. \tag{4}$$

This geometry describes a static, spherically symmetric black hole surrounded by a galactic halo, with the metric reducing to Schwarzschild as $V_c \to 0$. The horizon radius $r_h$ is defined as the largest positive root of the equation:

$$1 - \frac{2M}{r} + \frac{2V_c^2 r^2}{a^2 + r^2} = 0. \tag{5}$$

Expanding the solution to the above equation in terms of small $Vc$ we find the radius of the event horizon

$$r_h = 2M - \frac{16M^3 V_c^2}{a^2 + 4M^2} + \frac{128 M^5 V_c^4 \left(3a^2 + 4M^2\right)}{(a^2 + 4M^2)^3} + O(Vc^6). \tag{6}$$

In the following sections, we shall analyze the behavior of quasinormal modes for scalar and Dirac fields in this background, taking advantage of the analytic structure of the effective potential derived from (4).

## 3. Quasinormal Modes

Quasinormal modes (QNMs) characterize the damped oscillations of perturbation fields in the vicinity of black holes and other compact objects. They represent the response of a black hole to small external disturbances and are defined as solutions to the linearized perturbation equations subject to physically motivated boundary conditions.

For spherically symmetric spacetimes, the perturbation equations can typically be reduced to a Schrödinger-like wave equation of the form:

$$\frac{d^2 \Psi}{dr_*^2} + \left[\omega^2 - V(r)\right] \Psi = 0, \tag{7}$$

where $\Psi$ is the perturbation variable, $\omega$ is the (complex) quasinormal frequency, $V(r)$ is the effective potential depending on the type of field and background geometry, and $r_*$ is the tortoise coordinate defined via $dr_*/dr =$ $1/f(r)$, with $f(r)$ being the metric function.

Quasinormal boundary conditions require that the solution be purely ingoing at the event horizon and purely outgoing at spatial infinity (or at the cosmological horizon in asymptotically de Sitter spacetimes):

$$\Psi(r_*) \sim \begin{cases} e^{-i\omega r_*}, & r_* \to -\infty \quad \text{(near the horizon)}, \\ e^{+i\omega r_*}, & r_* \to +\infty \quad \text{(at infinity)}. \end{cases} \tag{8}$$

These boundary conditions lead to a discrete set of complex frequencies $\omega$, whose real part determines the oscillation frequency and imaginary part represents the decay rate of the mode.

To compute the QNMs, various numerical and semi-analytical methods are employed. One widely used approach is the WKB (Wentzel–Kramers–Brillouin) approximation, originally developed for quantum tunneling problems and adapted to black hole perturbations in [16–18]. In this method, the solution is matched across the turning points of the potential barrier. At $n$-th order, the WKB formula gives the complex frequencies approximately by:

$$\frac{iQ_0}{\sqrt{2Q_0''}} - \sum_{j=2}^{n} \Lambda_j = n + \frac{1}{2}, \quad n = 0, 1, 2, \ldots, \tag{9}$$

where $Q(r) = \omega^2 - V(r)$, and the quantities $\Lambda_j$ are higher-order WKB corrections expressed in terms of derivatives of the potential at its maximum.

To improve the convergence of the WKB approximation, especially for low multipole numbers $\ell$ or higher overtone numbers $n$, the use of Padé approximants has been proposed [19]. In this refinement, the WKB expansion of the function $\omega^2$ is expressed as a rational function using Padé techniques:

$$\omega^2 \approx P_n^m(\text{WKB series}) = \frac{A_0 + A_1 x + \cdots + A_m x^m}{1 + B_1 x + \cdots + B_n x^n}, \tag{10}$$

where $x$ is a formal parameter to label the WKB orders, and $P_n^m$ denotes the Padé approximant of order $(m, n)$. This resummation stabilizes the WKB series and leads to more accurate results, as confirmed by comparisons with numerical techniques such as direct integration or the continued-fraction (Leaver) method.

Throughout this work, we employ the sixth- and eighth-order WKB method with Padé approximants, $P_3^6$ and $P_3^8$, as a standard compromise between accuracy and computational complexity, which has proven reliable for a wide class of potentials with single-peak barrier structure.

*Scalar Field.* For a minimally coupled massless scalar field, the effective potential is:

$$V_{\text{scalar}}(r) = f(r) \left[ \frac{\ell(\ell+1)}{r^2} + \frac{f'(r)}{r} \right], \tag{11}$$

where $\ell = 0, 1, 2, \ldots$ is the multipole number.

*Electromagnetic Field.* For electromagnetic (Maxwell) perturbations, the effective potential is:

$$V_{\text{em}}(r) = f(r) \frac{\ell(\ell+1)}{r^2}, \quad \ell \geq 1. \tag{12}$$

*Dirac Field.* For massless Dirac perturbations, the effective potential is given in the form:

$$V_{\pm}(r) = W(r)^2 \pm \frac{dW(r)}{dr_*}, \tag{13}$$

where

$$W(r) = \left(\ell + \frac{1}{2}\right) \frac{\sqrt{f(r)}}{r}, \quad \ell = 1/2, 3/2, \ldots. \tag{14}$$

The two potentials $V_+$ and $V_-$ are isospectral, and it is sufficient to compute quasinormal modes using only one of them, typically $V_+$.

In the regime of high multipole numbers $\ell$, quasinormal frequencies can be found analytically. To this end, we express the position of the peak of the effective potential as a series expansion in the powers of the small parameter $V_c$. By solving equation $V'(r_{\max}) = 0$ and selecting the root that smoothly reduces to the Schwarzschild limit as $V_c \to 0$, we determine the coefficients of the expansion.

$$r_{max} = 3M - \frac{486 M^5 V_c^2}{(a^2 + 9M^2)^2} + O\left(\frac{1}{\kappa^2}, V_c^3\right) \tag{15}$$

Then using the first order WKB formula and expanding in small $1/\kappa$, and $V_c$, where $\kappa = \ell + 1/2$, we find the analytic form of the eikonal frequency:

$$\omega = \kappa \left( \frac{1}{3\sqrt{3}M} + \frac{3\sqrt{3}MV_c^2}{a^2 + 9M^2} \right) - \frac{iK}{3\sqrt{3}M} - \frac{3i\sqrt{3}MV_c^2 K \left(a^4 + 33a^2M^2 + 108M^4\right)}{(a^2 + 9M^2)^3} + O\left(\frac{1}{\kappa}, V_c^3\right). \quad (16)$$

Here we have

$$K = n + \frac{1}{2}.$$

Using the above expression for the eikonal modes we can immediately find the Lyapunov exponent for the unstable null geodesic and the orbital frequency at this orbit, because of the correspondence between null geodesics and eikonal quasinormal modes [20]. Indeed, according to this correspondence we have

$$\omega = \Omega_c \left(\ell + \frac{1}{2}\right) - i \left(n + \frac{1}{2}\right) |\lambda| + O(\ell^{-1}), \quad (17)$$

where $\Omega_c$ is the angular velocity and $\lambda$ the Lyapunov exponent of circular null geodesics. Notice, that while this correspondence does not hold in a number of cases [21–23], it does hold in our case because the centrifugal term has a canonical $f(r)\ell(\ell+1)r^{-2}$ form and the spacetime is asymptotically flat.

Tables 1–3 present the quasinormal frequencies for scalar, Dirac, and electromagnetic perturbations, computed using the sixth-order WKB method with Padé resummation. For all field types, the fundamental modes exhibit moderate deviations from their Schwarzschild counterparts as the dark matter parameter $V_c$ increases. These deviations are already noticeable for $\ell = 1$ and persist across different spins, indicating that the presence of a galactic halo can affect both the oscillation frequencies and the damping rates. While the effect is more pronounced at lower multipole numbers and generally diminishes with increasing $\ell$, the trend is not strictly monotonic and depends on the field spin and halo parameters. The agreement between the sixth- and eighth-order WKB results confirms the overall reliability of the numerical data. However, the case $\ell = 0$ suffers from reduced WKB accuracy and should be interpreted with caution.

**Table 1.** Quasinormal modes of the test scalar field ($s = 0$) for the black hole immersed in galactic halo ($M = 1$, $a = 10$) calculated using the WKB formula at different orders and the difference between them.

| $\ell$ | $Vc$ | **6th Order WKB** ($m = 3$) | **8th Order WKB** ($m = 3$) | **diff. (%)** |
|---|---|---|---|---|
| 0 | 0.01 | $0.111952 - 0.104578i$ | $0.111887 - 0.104718i$ | $0.101\%$ |
| 0 | 0.1 | $0.112192 - 0.104337i$ | $0.112193 - 0.104385i$ | $0.0315\%$ |
| 0 | 0.2 | $0.113154 - 0.103090i$ | $0.112938 - 0.103228i$ | $0.168\%$ |
| 0 | 0.3 | $0.115929 - 0.100899i$ | $0.113068 - 0.100770i$ | $1.86\%$ |
| 0 | 0.4 | $0.120050 - 0.097938i$ | $0.109501 - 0.097896i$ | $6.81\%$ |
| 0 | 0.5 | $0.126249 - 0.092557i$ | $0.102364 - 0.103979i$ | $16.9\%$ |
| 1 | 0.01 | $0.292940 - 0.097662i$ | $0.292942 - 0.097657i$ | $0.00161\%$ |
| 1 | 0.1 | $0.293826 - 0.097851i$ | $0.293828 - 0.097845i$ | $0.00209\%$ |
| 1 | 0.2 | $0.296502 - 0.098413i$ | $0.296505 - 0.098401i$ | $0.00402\%$ |
| 1 | 0.3 | $0.300943 - 0.099311i$ | $0.300947 - 0.099279i$ | $0.0100\%$ |
| 1 | 0.4 | $0.307130 - 0.100496i$ | $0.307129 - 0.100377i$ | $0.0367\%$ |
| 1 | 0.5 | $0.315003 - 0.101930i$ | $0.315040 - 0.101276i$ | $0.198\%$ |
| 2 | 0.01 | $0.483656 - 0.096761i$ | $0.483657 - 0.096761i$ | $0.00015\%$ |
| 2 | 0.1 | $0.484946 - 0.097001i$ | $0.484947 - 0.097001i$ | $0.00015\%$ |
| 2 | 0.2 | $0.488840 - 0.097723i$ | $0.488841 - 0.097723i$ | $0.00016\%$ |
| 2 | 0.3 | $0.495283 - 0.098908i$ | $0.495284 - 0.098908i$ | $0.00016\%$ |
| 2 | 0.4 | $0.504208 - 0.100530i$ | $0.504208 - 0.100528i$ | $0.00032\%$ |
| 2 | 0.5 | $0.515531 - 0.102552i$ | $0.515523 - 0.102546i$ | $0.00185\%$ |

**Table 2.** Quasinormal modes of the test Dirac field for the black hole immersed in galactic halo ($M = 1$, $a = 10$) calculated using the WKB formula at different orders and the difference between them.

| $\ell$ | $Vc$ | 6th Order WKB ($m = 3$) | 8th Order WKB ($m = 3$) | diff. (%) |
|---|---|---|---|---|
| 0.5 | 0.01 | $0.182642 - 0.096585i$ | $0.183165 - 0.096987i$ | 0.319% |
| 0.5 | 0.1 | $0.183053 - 0.096841i$ | $0.183587 - 0.097232i$ | 0.320% |
| 0.5 | 0.2 | $0.184289 - 0.097617i$ | $0.184852 - 0.097965i$ | 0.317% |
| 0.5 | 0.3 | $0.186328 - 0.098916i$ | $0.186916 - 0.099160i$ | 0.302% |
| 0.5 | 0.4 | $0.189162 - 0.100740i$ | $0.189703 - 0.100776i$ | 0.253% |
| 0.5 | 0.5 | $0.192806 - 0.103042i$ | $0.193097 - 0.102772i$ | 0.182% |
| 1.5 | 0.01 | $0.380043 - 0.096403i$ | $0.380047 - 0.096410i$ | 0.0021% |
| 1.5 | 0.1 | $0.380966 - 0.096665i$ | $0.380970 - 0.096672i$ | 0.0018% |
| 1.5 | 0.2 | $0.383750 - 0.097457i$ | $0.383752 - 0.097462i$ | 0.0012% |
| 1.5 | 0.3 | $0.388345 - 0.098766i$ | $0.388343 - 0.098767i$ | 0.0006% |
| 1.5 | 0.4 | $0.394690 - 0.100571i$ | $0.394681 - 0.100570i$ | 0.00223% |
| 1.5 | 0.5 | $0.402698 - 0.102848i$ | $0.402678 - 0.102847i$ | 0.00465% |
| 2.5 | 0.01 | $0.574108 - 0.096307i$ | $0.574108 - 0.096307i$ | 0.00006% |
| 2.5 | 0.1 | $0.575510 - 0.096572i$ | $0.575509 - 0.096572i$ | 0.00006% |
| 2.5 | 0.2 | $0.579735 - 0.097370i$ | $0.579735 - 0.097370i$ | 0.00008% |
| 2.5 | 0.3 | $0.586712 - 0.098689i$ | $0.586711 - 0.098689i$ | 0.00012% |
| 2.5 | 0.4 | $0.596344 - 0.100516i$ | $0.596342 - 0.100516i$ | 0.00027% |
| 2.5 | 0.5 | $0.608508 - 0.102830i$ | $0.608503 - 0.102829i$ | 0.00075% |

**Table 3.** Quasinormal modes of the test electromagnetic field for the black hole immersed in galactic halo ($M = 1$, $a = 10$) calculated using the WKB formula at different orders and the difference between them.

| $\ell$ | $Vc$ | 6th Order WKB ($m = 3$) | 8th Order WKB ($m = 3$) | diff. (%) |
|---|---|---|---|---|
| 1 | 0.01 | $0.248257 - 0.092482i$ | $0.248272 - 0.092491i$ | 0.0065% |
| 1 | 0.1 | $0.248850 - 0.092723i$ | $0.248865 - 0.092732i$ | 0.0065% |
| 1 | 0.2 | $0.250639 - 0.093449i$ | $0.250655 - 0.093455i$ | 0.0064% |
| 1 | 0.3 | $0.253597 - 0.094642i$ | $0.253615 - 0.094642i$ | 0.0065% |
| 1 | 0.4 | $0.257692 - 0.096281i$ | $0.257715 - 0.096260i$ | 0.0113% |
| 1 | 0.5 | $0.262889 - 0.098344i$ | $0.262939 - 0.098249i$ | 0.0384% |
| 2 | 0.01 | $0.457606 - 0.095007i$ | $0.457607 - 0.095007i$ | 0.00016% |
| 2 | 0.1 | $0.458716 - 0.095265i$ | $0.458717 - 0.095265i$ | 0.00016% |
| 2 | 0.2 | $0.462063 - 0.096045i$ | $0.462064 - 0.096045i$ | 0.00016% |
| 2 | 0.3 | $0.467589 - 0.097333i$ | $0.467589 - 0.097333i$ | 0.00017% |
| 2 | 0.4 | $0.475218 - 0.099115i$ | $0.475218 - 0.099114i$ | 0.00019% |
| 2 | 0.5 | $0.484854 - 0.101370i$ | $0.484852 - 0.101367i$ | 0.00066% |
| 3 | 0.01 | $0.656915 - 0.095619i$ | $0.656915 - 0.095619i$ | 0.00001% |
| 3 | 0.1 | $0.658516 - 0.095881i$ | $0.658516 - 0.095881i$ | 0.00001% |
| 3 | 0.2 | $0.663344 - 0.096673i$ | $0.663344 - 0.096673i$ | 0.00002% |
| 3 | 0.3 | $0.671314 - 0.097982i$ | $0.671314 - 0.097982i$ | 0.00001% |
| 3 | 0.4 | $0.682317 - 0.099796i$ | $0.682317 - 0.099796i$ | 0.00001% |
| 3 | 0.5 | $0.696211 - 0.102095i$ | $0.696210 - 0.102095i$ | 0.00005% |

A more important observation is that, as the halo parameter $a$ increases, the quasinormal frequencies rapidly approach their Schwarzschild values. In fact, for $a \gg M$, the frequencies become practically indistinguishable from those of a Schwarzschild black hole in vacuum (see Table 4). We therefore conclude that, in order to produce any observable deviation, the dark matter halo must be extraordinarily compact and dense. Similar conclusions were drawn in Refs. [24, 25] for other halo models.

**Table 4.** Quasinormal modes of the electromagnetic perturbations for the black hole immersed in galactic halo ($M = 1$, $V_c = 0.3$) calculated using the WKB formula at different orders and the difference between them.

| $\ell$ | $a$ | 6th Order WKB ($m = 3$) | 8th Order WKB ($m = 3$) | diff. (%) |
|---|---|---|---|---|
| 1 | 1 | $0.302742 - 0.125183i$ | $0.302778 - 0.125289i$ | $0.0340\%$ |
| 1 | 2 | $0.288294 - 0.117944i$ | $0.288068 - 0.118086i$ | $0.0857\%$ |
| 1 | 4 | $0.269882 - 0.105216i$ | $0.270146 - 0.105183i$ | $0.0919\%$ |
| 1 | 6 | $0.261133 - 0.098936i$ | $0.261210 - 0.099003i$ | $0.0367\%$ |
| 1 | 8 | $0.256279 - 0.095959i$ | $0.256315 - 0.095974i$ | $0.0140\%$ |
| 1 | 10 | $0.253597 - 0.094642i$ | $0.253615 - 0.094642i$ | $0.0065\%$ |
| 1 | 15 | $0.250707 - 0.093432i$ | $0.250719 - 0.093443i$ | $0.0059\%$ |
| 1 | 20 | $0.249650 - 0.093017i$ | $0.249664 - 0.093027i$ | $0.0066\%$ |
| 1 | 30 | $0.248879 - 0.092719i$ | $0.248894 - 0.092728i$ | $0.0066\%$ |
| 2 | 1 | $0.565892 - 0.129496i$ | $0.565896 - 0.129497i$ | $0.00083\%$ |
| 2 | 2 | $0.537446 - 0.121843i$ | $0.537432 - 0.121858i$ | $0.00356\%$ |
| 2 | 4 | $0.499617 - 0.108039i$ | $0.499619 - 0.108029i$ | $0.00204\%$ |
| 2 | 6 | $0.481526 - 0.101479i$ | $0.481528 - 0.101484i$ | $0.00123\%$ |
| 2 | 8 | $0.472475 - 0.098665i$ | $0.472477 - 0.098665i$ | $0.00040\%$ |
| 2 | 10 | $0.467589 - 0.097333i$ | $0.467589 - 0.097333i$ | $0.00017\%$ |
| 2 | 15 | $0.462259 - 0.096029i$ | $0.462259 - 0.096029i$ | $0.00014\%$ |
| 2 | 20 | $0.460265 - 0.095578i$ | $0.460265 - 0.095578i$ | $0.00016\%$ |
| 2 | 30 | $0.458796 - 0.095259i$ | $0.458797 - 0.095259i$ | $0.00016\%$ |
| 3 | 1 | $0.814993 - 0.130529i$ | $0.814994 - 0.130529i$ | $0.00008\%$ |
| 3 | 2 | $0.773664 - 0.122778i$ | $0.773662 - 0.122780i$ | $0.00043\%$ |
| 3 | 4 | $0.717834 - 0.108688i$ | $0.717834 - 0.108687i$ | $0.00014\%$ |
| 3 | 6 | $0.691370 - 0.102128i$ | $0.691370 - 0.102129i$ | $0.00010\%$ |
| 3 | 8 | $0.678334 - 0.099326i$ | $0.678334 - 0.099325i$ | $0.00004\%$ |
| 3 | 10 | $0.671314 - 0.097982i$ | $0.671314 - 0.097982i$ | $0.00001\%$ |
| 3 | 15 | $0.663640 - 0.096657i$ | $0.663640 - 0.096657i$ | $0\%$ |
| 3 | 20 | $0.660761 - 0.096199i$ | $0.660761 - 0.096199i$ | $0.00001\%$ |
| 3 | 30 | $0.658638 - 0.095874i$ | $0.658638 - 0.095874i$ | $0.00001\%$ |

Although the quasinormal spectra exhibit some sensitivity to halo parameters in the extreme regime of compact and dense halos, these conditions are not representative of realistic astrophysical settings. This implies that gravitational wave observations of black hole ringdowns are unlikely to capture halo-induced modifications unless highly exotic or nonstandard dark matter configurations are present. Thus, deviations from the Schwarzschild QNM spectrum remain a clean probe of fundamental modifications to gravity, rather than dark matter environment effects.

## 4. Beyond Eikonal Expansion

The above eikonal formula can be expanded to higher orders using the higher orders of the WKB series. The whole approach is described in [26] and applied in numerous papers [27–31], so here we will only summarize the results.

*Scalar field.* Using the series expansion in terms of the inverse multipole number [26], for the scalar field we find the expansion for the position of a maximum of the effective potential,

$$r_{\max} = 3M - \frac{M}{3\kappa^2} + V_c^2 \left( \frac{18a^2 M^3 \left(a^2 - 15M^2\right)}{\kappa^2 \left(a^2 + 9M^2\right)^3} - \frac{486M^5}{\left(a^2 + 9M^2\right)^2} \right) + \mathcal{O}\left(V_c^4, \frac{1}{\kappa^4}\right). \quad (18)$$

Then, using the WKB formula, we obtain the expression for the frequency

$$\begin{aligned}
\omega = &-\frac{iK\left(940K^2+313\right)}{46656\sqrt{3}M\kappa^2} + \frac{29-60K^2}{1296\sqrt{3}M\kappa} + \frac{\kappa}{3\sqrt{3}M} - \frac{iK}{3\sqrt{3}M} \\
&+ V_c^2\Bigg(\frac{3\sqrt{3}M\kappa}{a^2+9M^2} - \frac{3i\sqrt{3}MK(a^4+33a^2M^2+108M^4)}{(a^2+9M^2)^3} \\
&- \frac{M\left(a^8(60K^2-101) - 2a^6M^2(780K^2+1549)\right)}{16\sqrt{3}\kappa(a^2+9M^2)^5} \\
&+ \frac{M\left(18a^4M^4(1380K^2-1153) + 162a^2M^6(852K^2-617) + 729M^8(300K^2-253)\right)}{16\sqrt{3}\kappa(a^2+9M^2)^5} \\
&- \frac{iMK\left(a^{12}(4700K^2-15571) - 45a^{10}M^2(9836K^2+6809)\right)}{1728\sqrt{3}\kappa^2(a^2+9M^2)^7} \\
&+ \frac{iMK\left(324a^8M^4(19660K^2+33757) + 1458a^6M^6(32660K^2+19391)\right)}{1728\sqrt{3}\kappa^2(a^2+9M^2)^7} \\
&+ \frac{iMK\left(6561a^4M^8(37340K^2+4541) + 59049a^2M^{10}(14260K^2+4687)\right)}{1728\sqrt{3}\kappa^2(a^2+9M^2)^7} \\
&+ \frac{iMK \cdot 1062882M^{12}(940K^2+313)}{1728\sqrt{3}\kappa^2(a^2+9M^2)^7}\Bigg) + \mathcal{O}\left(V_c^4, \frac{1}{\kappa^3}\right)
\end{aligned} \quad (19)$$

Here we used the following designations: $\kappa \equiv \ell + 1/2$, $K \equiv n + 1/2$.

*Dirac field.* For Dirac field perturbations, in a similar way we have

$$\begin{aligned}
r_{\max} = &\frac{11M}{16\sqrt{3}\kappa^3} - \frac{\sqrt{3}M}{2\kappa} + 3M \\
&+ V_c^2\Bigg(\frac{81a^2M^5(5a^2-9M^2)}{\kappa^2(a^2+9M^2)^4} - \frac{486M^5}{(a^2+9M^2)^2} \\
&+ \frac{27\sqrt{3}M^3(3M^2-a^2)}{2\kappa(a^2+9M^2)^2} + \frac{27\sqrt{3}M^3\left(5a^8+58a^6M^2+3672a^4M^4\right)}{16\kappa^3(a^2+9M^2)^5} \\
&+ \frac{27\sqrt{3}M^3\left(12150a^2M^6+8019M^8\right)}{16\kappa^3(a^2+9M^2)^5}\Bigg) + \mathcal{O}\left(V_c^4, \frac{1}{\kappa^4}\right),
\end{aligned} \quad (20)$$

for the position of the peak and

$$\begin{aligned}
\omega = &\frac{iK(119-940K^2)}{46656\sqrt{3}M\kappa^2} - \frac{60K^2+7}{1296\sqrt{3}M\kappa} + \frac{\kappa}{3\sqrt{3}M} - \frac{iK}{3\sqrt{3}M} \\
&+ V_c^2\Bigg(\frac{3\sqrt{3}M\kappa}{a^2+9M^2} - \frac{3i\sqrt{3}MK(a^4+33a^2M^2+108M^4)}{(a^2+9M^2)^3} \\
&- \frac{M\left(a^8(60K^2+7) - 2a^6M^2(780K^2+163)\right)}{16\sqrt{3}\kappa(a^2+9M^2)^5} \\
&+ \frac{M\left(90a^4M^4(276K^2+61) + 162a^2M^6(852K^2+85) + 3645M^8(60K^2+7)\right)}{16\sqrt{3}\kappa(a^2+9M^2)^5} \\
&- \frac{iMK\left(a^{12}(4700K^2+413) - 45a^{10}M^2(9836K^2+6377)\right)}{1728\sqrt{3}\kappa^2(a^2+9M^2)^7} \\
&+ \frac{iMK\left(324a^8M^4(19660K^2+20797) + 1458a^6M^6(32660K^2-6097)\right)}{1728\sqrt{3}\kappa^2(a^2+9M^2)^7} \\
&+ \frac{iMK\left(32805a^4M^8(7468K^2-2807) + 59049a^2M^{10}(14260K^2-1793)\right)}{1728\sqrt{3}\kappa^2(a^2+9M^2)^7} \\
&+ \frac{iMK \cdot 1062882M^{12}(940K^2-119)}{1728\sqrt{3}\kappa^2(a^2+9M^2)^7}\Bigg) + \mathcal{O}\left(V_c^4, \frac{1}{\kappa^3}\right)
\end{aligned} \quad (21)$$

for the frequency.

*Maxwell field.* In the same way for electromagnetic perturbations we have the following expressions for the position of the maximum of the effective potential and the quasinormal modes

$$r_{\max} = 3M - \frac{486 M^5 V_c^2}{(a^2+9M^2)^2} + \mathcal{O}\left(V_c^4, \frac{1}{\kappa^4}\right), \tag{22}$$

$$\begin{aligned}
\omega = &-\frac{5iK(188K^2-283)}{46656\sqrt{3}M\kappa^2} - \frac{5(12K^2+23)}{1296\sqrt{3}M\kappa} + \frac{\kappa}{3\sqrt{3}M} - \frac{iK}{3\sqrt{3}M} \\
&+ V_c^2 \Bigg( \frac{3\sqrt{3}M\kappa}{a^2+9M^2} - \frac{3i\sqrt{3}MK(a^4+33a^2M^2+108M^4)}{(a^2+9M^2)^3} \\
&- \frac{M\left(a^8(60K^2+43) + 10a^6 M^2(151-156K^2)\right)}{16\sqrt{3}\kappa(a^2+9M^2)^5} \\
&+ \frac{M\left(18a^4 M^4(1380K^2+1871) + 162a^2 M^6(852K^2+1111) + 729 M^8(300K^2+467)\right)}{16\sqrt{3}\kappa(a^2+9M^2)^5} \\
&- \frac{iMK\left(25a^{12}(188K^2-139) - 45a^{10}M^2(9836K^2+5081)\right)}{1728\sqrt{3}\kappa^2(a^2+9M^2)^7} \\
&+ \frac{iMK\left(1620a^8 M^4(3932K^2+3641) + 1458a^6 M^6(32660K^2-41089)\right)}{1728\sqrt{3}\kappa^2(a^2+9M^2)^7} \\
&+ \frac{iMK\left(6561a^4 M^8(37340K^2-80131) + 59049a^2 M^{10}(14260K^2-21233)\right)}{1728\sqrt{3}\kappa^2(a^2+9M^2)^7} \\
&+ \frac{iMK \cdot 5314410 M^{12}(188K^2-283)}{1728\sqrt{3}\kappa^2(a^2+9M^2)^7} \Bigg) + \mathcal{O}\left(V_c^4, \frac{1}{\kappa^3}\right).
\end{aligned} \tag{23}$$

Notice that in the eikonal regime all the expression for frequencies and the position of the peak do not depend on the spin of the field. The difference in spin appear only at the orders beyond the eikonal limit.

## 5. Unruh Temperature

Verlinde's emergent gravity framework [32] interprets gravity not as a fundamental force, but as an entropic phenomenon arising from microscopic degrees of freedom encoded holographically on screens. In this approach, the gravitational force experienced by a test particle is associated with changes in entropy when the particle moves relative to such a screen. A key ingredient in this formulation is the Unruh temperature, the effective temperature measured by an accelerated observer near the screen, which plays a role analogous to temperature in thermodynamics. In the context of static, spherically symmetric black hole spacetimes, the Unruh temperature can be expressed in terms of the redshifted surface gravity or, more generally, through covariant quantities involving the gravitational potential and the timelike Killing vector. This temperature determines the local acceleration and, ultimately, the entropic force that reproduces Newtonian gravity in the appropriate limit. In the setup, the Unruh temperature was calculated for various configurations [24, 32, 33] and here the Unruh temperature is computed for black hole solutions surrounded by the dark matter distributions of the galactic halo, thus extending Verlinde's framework beyond vacuum spacetimes.

Using the the gravitational potential $\varphi$, and a timelike Killing vector field $\xi^\alpha$,

$$\varphi = \frac{1}{2}\log\left(-g_{\alpha\beta}\xi^\alpha\xi^\beta\right), \tag{24}$$

one can find the local acceleration as follows

$$a^\alpha = -g^{\alpha\beta}\nabla_\beta\varphi. \tag{25}$$

Then, the Unruh temperature is [32, 34]

$$T_{Unruh} = \frac{\hbar}{2\pi}e^\varphi\, n^\alpha \nabla_\alpha \varphi, \tag{26}$$

which further transform to the following form

$$T_{Unruh} = \frac{\hbar}{2\pi} \frac{e^{\varphi}}{\sqrt{g^{\alpha\beta} \partial_{\alpha}\varphi \, \partial_{\beta}\varphi}}, \tag{27}$$

Finally, for the spherically symmetric black hole ith the metric function $f(r)$ the Unruh temperature as measured by a static observer at radius $r$, has the form

$$T_{\text{Unruh}}(r) = \hbar \frac{f'(r)}{4\pi}, \tag{28}$$

which, using the units $\hbar = 1$ can be written explicitly in the following way

$$T_{\text{Unruh}}(r) = \frac{a^4 M + 2a^2 r^2 \left(M + r\mathbf{V}\mathbf{c}^2\right) + Mr^4}{2\pi r^2 \left(a^2 + r^2\right)^2}. \tag{29}$$

Figure 1 illustrates the Unruh-Verlinde temperature as perceived by a static observer at a fixed radial coordinate for various values of the halo parameters $V_c$ and $a$. The results demonstrate that this local temperature is highly sensitive to the compactness and density of the dark matter halo. Specifically, smaller values of $a$ and larger $V_c$—corresponding to denser and more centrally concentrated halos—lead to substantial increases in the Unruh-Verlinde temperature. This enhancement reflects the stronger proper acceleration required to maintain a static position against the increased gravitational pull due to the dark matter distribution. In contrast, for realistic halo parameters with large $a$, the temperature profile approaches that of the Schwarzschild black hole, reaffirming that galactic halos must be extremely compact to induce appreciable local thermodynamic deviations.

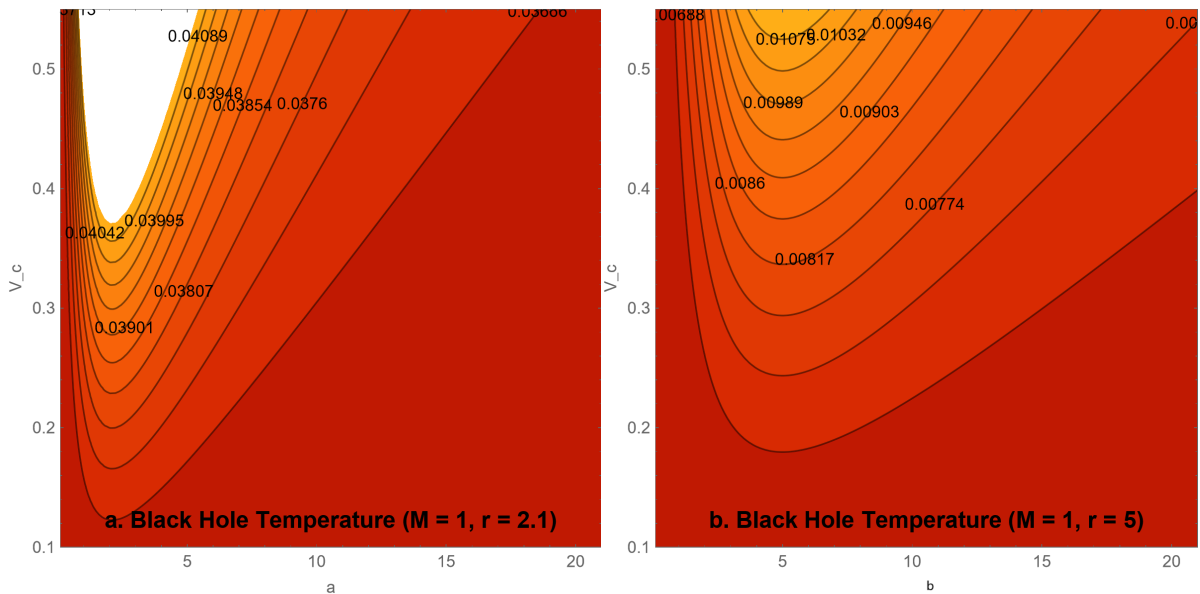

**Figure 1.** Unruh temperature at fixed radial coordinates: $r = 2.1$ in the left panel (**a**) and $r = 5$ in the right panel (**b**), as a function of $V_c$ and $a$; $M = 1$.

## 6. Conclusions

We have studied quasinormal modes and Unruh temperature for test scalar, electromagnetic, and Dirac fields in the background of a static black hole immersed in a spherically symmetric dark matter halo. The spacetime is sourced by a physically motivated matter distribution producing a flat galactic rotation curve at large distances.

Using the sixth-order WKB method with Padé resummation, we computed the fundamental quasinormal frequencies and verified their consistency with higher-order approximations. We also derived analytic expressions in the eikonal and beyond-eikonal regimes. Our results show that the presence of the dark matter halo leads to small but measurable deviations in the quasinormal spectra only when the halo is extremely compact and dense.

However, for astrophysically realistic halo parameters, where the mass and extent of the halo are much greater than the black hole's horizon scale, the influence of the halo on quasinormal modes becomes negligible. The frequencies quickly converge to their Schwarzschild values, and the Unruh temperature remains effectively unaffected. These findings are in agreement with earlier results for other halo models [24, 25] and confirm that quasinormal ringing remains a robust observable for testing strong-field gravity, even in the presence of diffuse dark matter environments.

**Funding**

This research received no external funding.

**Institutional Review Board Statement**

Not applicable.

**Informed Consent Statement**

Not applicable.

**Data Availability Statement**

Not applicable.

**Acknowledgments**

The author acknowledges the University of Seville for their support through the Plan-US of aid to Ukraine.

**Conflicts of Interest**

The author declares no conflict of interest.

**References**

1. Kokkotas, K.D.; Schmidt, B.G. Quasinormal modes of stars and black holes. *Living Rev. Rel.* **1999**, *2*, 1–72. https://doi.org/10.12942/lrr-1999-2.
2. Nollert, H.-P. TOPICAL REVIEW: Quasinormal modes: The characteristic 'sound' of black holes and neutron stars. *Class. Quant. Grav.* **1999**, *16*, R159–R216. https://doi.org/10.1088/0264-9381/16/12/201.
3. Bolokhov, S.V.; Skvortsova, M. Review of analytic results on quasinormal modes of black holes. *arXiv* **2025**, arXiv:2504.05014.
4. Pezzella, L.; Destounis, K.; Maselli, A.; et al. Quasinormal modes of black holes embedded in halos of matter. *Phys. Rev. D* **2025**, *111*, 064026. https://doi.org/10.1103/PhysRevD.111.064026.
5. Liu, D.; Yang, Y.; Long, Z. Probing the black holes in a dark matter halo of M87 using gravitational wave echoes. *Eur. Phys. J. C* **2024**, *84*, 871. https://doi.org/10.1140/epjc/s10052-024-13255-x.
6. Liu, Y.; Mu, B.; Tao, J.; et al. Quasinormal modes of Schwarzschild-like black hole surrounded by the pseudo-isothermal dark matter halo. *Nucl. Phys. B* **2025**, *1010*, 116787. https://doi.org/10.1016/j.nuclphysb.2024.116787.
7. Mollicone, A.; Destounis, K. Superradiance of charged black holes embedded in dark matter halos. *Phys. Rev. D* **2025**, *111*, 024017. https://doi.org/10.1103/PhysRevD.111.024017.
8. Chen, Ru.; Javed, F.; Mustafa, D.G.; et al. Dual effect of string cloud and dark matter halos on particle motions, shadows and epicyclic oscillations around Schwarzschild black holes. *JHEAp* **2024**, *44*, 172–186. https://doi.org/10.1016/j.jheap.2024.09.010.
9. Jha, S.K. Shadow, ISCO, quasinormal modes, Hawking spectrum, weak gravitational lensing, and parameter estimation of a Schwarzschild black hole surrounded by a Dehnen type dark matter halo. *JCAP* **2025**, *3*, 54. https://doi.org/10.1088/1475-7516/2025/03/054.
10. Bécar, R.; González, P.A.; Papantonopoulos, E.; et al. Massive scalar field perturbations of black holes immersed in Chaplygin-like dark fluid. *JCAP* **2024**, *6*, 61. https://doi.org/10.1088/1475-7516/2024/06/061.
11. Liu, D.; Yang, Y.; Long, Z. Probing black holes in a dark matter spike of M87 using quasinormal modes. *Eur. Phys. J. C* **2024**, *84*, 731. https://doi.org/10.1140/epjc/s10052-024-13096-8.
12. Hamil, B.; Al-Badawi, A.; Lütfüoğlu, B.C. Geodesics and scalar perturbations of Schwarzschild black holes embedded in a Dehnen-type dark matter halo with quintessence. *arXiv* **2025**, arXiv: 2505.18611.
13. Lobo, F.S.N.; Ramos, J.A.A.; Rodrigues, M.E. Supermassive black hole in NGC 4649 (M60) with a dark matter halo: Impact on shadow measurements and thermodynamic properties. *arXiv* **2025**, arXiv:2505.03661.
14. Ma, S.; Wang, R.; Deng, Ji.; et al. Euler–Heisenberg black hole surrounded by perfect fluid dark matter. *Eur. Phys. J. C* **2024**, *84*, 595. https://doi.org/10.1140/epjc/s10052-024-12914-3.
15. Shen, J.; Gebhardt, K. The Supermassive Black Hole and Dark Matter Halo of NGC 4649 (M60). *Astrophys. J.* **2010**, *711*, 484–494. https://doi.org/10.1088/0004-637X/711/1/484.
16. Schutz, B.F.; Will, C.M. Black hole normal modes: A semianalytic approach. *Astrophys. J. Lett.* **1985**, *291*, L33–L36. https://doi.org/10.1086/184453.

17. Iyer, S.; Will, C.M. Black Hole Normal Modes: A WKB Approach. 1. Foundations and Application of a Higher Order WKB Analysis of Potential Barrier Scattering. *Phys. Rev. D* **1987**, *35*, 3621. https://doi.org/10.1103/PhysRevD.35.3621.
18. Konoplya, R.A. Quasinormal behavior of the d-dimensional Schwarzschild black hole and higher order WKB approach. *Phys. Rev. D* **2003**, *68*, 24018. https://doi.org/10.1103/PhysRevD.68.024018.
19. Matyjasek, J.; Opala, M. Quasinormal modes of black holes. The improved semianalytic approach. *Phys. Rev. D* **2017**, *96*, 024011. https://doi.org/10.1103/PhysRevD.96.024011.
20. Cardoso, V.; Miranda, A.S.; Berti, E.; et al. Geodesic stability, Lyapunov exponents and quasinormal modes. *Phys. Rev. D* **2009**, *79*, 064016. https://doi.org/10.1103/PhysRevD.79.064016.
21. Khanna, G.; Price, R.H. Black Hole Ringing, Quasinormal Modes, and Light Rings. *Phys. Rev. D* **2017**, *95*, 081501. https://doi.org/10.1103/PhysRevD.95.081501.
22. Konoplya, R.A. Further clarification on quasinormal modes/circular null geodesics correspondence. *Phys. Lett. B* **2023**, *838*, 137674. https://doi.org/10.1016/j.physletb.2023.137674.
23. Bolokhov, S.V. Black holes in Starobinsky-Bel-Robinson Gravity and the breakdown of quasinormal modes/null geodesics correspondence. *Phys. Lett. B* **2024**, *856*, 138879. https://doi.org/10.1016/j.physletb.2024.138879.
24. Konoplya, R.A. Black holes in galactic centers: Quasinormal ringing, grey-body factors and Unruh temperature. *Phys. Lett. B* **2021**, *823*, 136734. https://doi.org/10.1016/j.physletb.2021.136734.
25. Konoplya, R.A.; Zhidenko, A. Solutions of the Einstein Equations for a Black Hole Surrounded by a Galactic Halo. *Astrophys. J.* **2022**, *933*, 166. https://doi.org/10.3847/1538-4357/ac76bc.
26. Konoplya, R.A.; Zhidenko, A. Analytic expressions for quasinormal modes and grey-body factors in the eikonal limit and beyond. *Class. Quant. Grav.* **2023**, *40*, 245005. https://doi.org/10.1088/1361-6382/ad0a52.
27. Bolokhov, S.V. Long-lived quasinormal modes and overtones' behavior of holonomy-corrected black holes. *Phys. Rev. D* **2024**, *110*, 024010. https://doi.org/10.1103/PhysRevD.110.024010.
28. Dubinsky, A. Analytic expressions for quasinormal modes of the general parametrized spherically symmetric black holes and the Hod's proposal. *Phys. Lett. B* **2025**, *861*, 139251. https://doi.org/10.1016/j.physletb.2025.139251.
29. Malik, Z. Analytical QNMs of fields of various spin in the Hayward spacetime. *EPL* **2024**, *147*, 69001. https://doi.org/10.1209/0295-5075/ad7885.
30. Malik, Z. Quasinormal Modes of Dilaton Black Holes: Analytic Approximations. *Int. J. Theor. Phys.* **2024**, *63*, 128. https://doi.org/10.1007/s10773-024-05660-5.
31. Dubinsky, A.; Zinhailo, A.F. Analytic expressions for grey-body factors of the general parametrized spherically symmetric black holes. *EPL* **2025**, *149*, 69004. https://doi.org/10.1209/0295-5075/adbc17.
32. Verlinde, E.P. On the Origin of Gravity and the Laws of Newton. *JHEP* **2011**, *4*, 029. https://doi.org/10.1007/JHEP04(2011)029.
33. Konoplya, R.A. Entropic force, holography and thermodynamics for static space-times. *Eur. Phys. J. C* **2010**, *69*, 555–562. https://doi.org/10.1140/epjc/s10052-010-1424-1.
34. Unruh, W.G. Notes on black hole evaporation. *Phys. Rev. D* **1976**, *14*, 870. https://doi.org/10.1103/PhysRevD.14.870.